# Enhanced boiling heat transfer using conducting–insulating microcavity surfaces in an electric field: A lattice Boltzmann study ⊘

Fanming Cai (蔡凡茗)  ; Zhaomiao Liu (刘赵淼)   ; Nan Zheng (郑楠)  ; Yan Pang (逄燕)

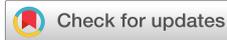

View
Online

Export
Citation

CrossMark

---

**Articles You May Be Interested In**

A functional integral formalism for quantum spin systems

*J. Math. Phys.* (July 2008)

Modes selection in polymer mixtures undergoing phase separation by photochemical reactions

*Chaos* (June 1999)

Spreading of a surfactant monolayer on a thin liquid film: Onset and evolution of digitated structures

*Chaos* (March 1999)

**Physics of Fluids**

AIP Publishing

# Enhanced boiling heat transfer using conducting–insulating microcavity surfaces in an electric field: A lattice Boltzmann study

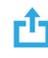

Fanming Cai (蔡凡茗), Zhaomiao Liu (刘赵淼),a) Nan Zheng (郑楠), and Yan Pang (逄燕)

### AFFILIATIONS

Faculty of Materials and Manufacturing, Beijing University of Technology, Beijing, China

a)Author to whom correspondence should be addressed: lzm@bjut.edu.cn

### ABSTRACT

The field trap effect on the microcavity surface under the action of an electric field is not conducive to boiling heat transfer. This numerical study found that using conducting–insulating microcavity surfaces in an electric field removes the field trap effect, increasing the critical heat flux by more than 200%. Bubble behavior and heat transfer mechanisms on heating surfaces were further explored. The results show that a large electrical force can be generated at the junction of the conducting and insulating surfaces under the action of the electric field, which drives the bubbles in the cavity to depart quickly from the heating surface and avoids the formation of a vapor block. As the electric field intensity increases, the contact line produces pinning, which facilitates the formation of multiple continuously open vapor–liquid separation paths on the heating surface, resulting in a significant enhancement of the boiling heat transfer performance. Finally, a modified correlation equation is proposed to predict the critical heat flux under non-uniform electric field.

Published under an exclusive license by AIP Publishing. https://doi.org/10.1063/5.0171247

## I. INTRODUCTION

Boiling heat transfer utilizes the latent heat of vapor–liquid phase change and bubble-induced micro-convection, which allows heat transfer coefficients to be increased by orders of magnitude compared with traditional single-phase convection and thus is widely used in the cooling process of a variety of microelectronic devices characterized by high thermal loads.[1,2] However, when the heat transfer power exceeds the limit [i.e., critical heat flux (CHF)], the heating surface is rapidly covered by a vapor film that causes the surface temperature to surge, which may lead to severe damage to the equipment. Therefore, improving the CHF of the heating surface is important for developing the electronics industry.

Existing schemes for enhancing critical heat flux can be categorized into passive and active techniques. Typical strategies for passive techniques are to utilize surface microstructures (e.g., microcavities, micropillars, and porous media) to decrease the bubble departure size,[3,4] increase the bubble departure frequency,[5,6] and enhance capillary wicking[7,8] to improve the CHF. However, under microgravity conditions, due to the disappearance of buoyancy, achieving the effective removal of bubbles only by microstructure is difficult, which is likely to cause vapor film formation on the heating surface and thus reduce the CHF.[9] Introducing an external force field instead of the buoyancy effect is particularly important in this situation. Electric field as an active control scheme has the advantages of low power consumption, non-contact, and high stability, which can effectively improve boiling heat transfer performance.[10] Previous studies have shown that electrical forces can avoid bubbles merging at high superheat levels by squeezing the vapor–liquid interface, which delays CHF occurrence.[11,12] Under the influence of an electric field, the boiling process can produce bubble sizes close to those of normal gravity conditions, even in a microgravity environment.[13]

To combine the respective advantages of electric fields and microstructured surfaces, researchers in recent years have focused on the effect of the coupled behavior of both on the performance of boiling heat transfer. Garivalis et al.[14] experimentally demonstrated that electric fields and microstructured surfaces can enhance boiling heat transfer synergistically. Even more significant enhancement of CHF can be achieved in a microgravity environment. However, Quan[15] and Liu et al.[16] suggested that coupling electric field and microstructured surfaces would inhibit the boiling heat transfer. The above-mentioned differences may come from the different surface structure parameters. When the height (depth) of the micropillar (microcavity) used is large, the region of high electric field intensity formed at its top will inhibit the growth and departure of bubbles at the root (i.e., the field trap effect), thus deteriorating the boiling performance. Recently, several methods have been proposed to mitigate the field trap effect. The researchers set up hydrophobic spots[17] or holes[18] in the high electric field intensity region at the top of the microstructures to promote bubble nucleation and departure. However, it should be emphasized that bubbles are still generated at the root of the microstructure when the superheat is sufficiently large, and thus, the inhibition of boiling heat transfer by the field trap effect is not eliminated. Designing novel heating surfaces is urgently needed to realize the optimal heat transfer effect of the combination of electric field and microstructured surface.

This study proposes a conducting–insulating microcavity surface that can eliminate the field trap effect through lattice Boltzmann simulations. The boiling heat transfer mechanism of conducting–insulating microcavity surfaces under the influence of an electric field is further revealed by exploring the thermodynamic relationship between the bubble behavior and the thermal response of the heater bottom. The study results can provide theoretical and applied technical guidance for the structural design of novel heating surfaces.

## II. METHODOLOGIES
### A. Model description

Based on the previous studies,[15,16] it is known that the field trap effect comes from the high electric field intensity generated at the corners of the microstructures. Therefore, the field trap effect may be eliminated after the corner structure is insulated. To verify the feasibility of the method and investigate the corresponding boiling mechanism, a conducting–insulating surface is designed and a numerical model of vapor–liquid two-phase flow and phase change heat transfer is developed in this study, as shown in Fig. 1. The bottom of the computational domain is a heating surface with microcavities of height $H_0$, width $W_0$, and depth $D_0$. The central regions at the bottom and top of the microcavities have a conducting block of width $W_c$, and the rest of the surfaces are insulated. When the height from the top of the microcavity is greater than $H_e$, the system maintains a constant high electric potential $U_0$, and the electric potential of the conducting block is zero (electric field intensity $E_0 = U_0/H_e$). The high thermal load $T_b$ is applied at the bottom of the computational domain, while the top maintains the saturation temperature $T_{sat}$. The thermal response of the solid–liquid interface needs to be obtained by numerical calculations and is not predetermined.

### B. The pseudopotential lattice Boltzmann method

The governing equation for obtaining the vapor–liquid two-phase flow based on the multi-relaxation pseudo-potential lattice Boltzmann method can be formulated as follows:[19]

$$f_\alpha(\boldsymbol{x} + \boldsymbol{e}_\alpha \Delta t, t + \Delta t) = f_\alpha(\boldsymbol{x}, t) - (\mathbf{M}^{-1} \mathbf{\Lambda} \mathbf{M})_{\alpha\beta}$$
$$\times [f_\alpha(\boldsymbol{x}, t) - f_\alpha^{eq}(\boldsymbol{x}, t)] + \Delta t F'_\alpha(\boldsymbol{x}, t),$$
$$\rho(\boldsymbol{x}, t) = \sum_\alpha^8 f_\alpha(\boldsymbol{x}, t), \tag{1}$$
$$\rho\boldsymbol{u}(\boldsymbol{x}, t) = \sum_\alpha^8 f_\alpha(\boldsymbol{x}, t)\boldsymbol{e}_\alpha + \frac{\boldsymbol{F}(\boldsymbol{x}, t)\Delta t}{2},$$

where $f_\alpha$ is the distribution function related to density, and $\boldsymbol{x}$ and $t$ denote the spatial location and time of the distribution function, respectively. $\Delta t = 1$ is the time step, and $\boldsymbol{e}_\alpha$ (where $\alpha$ is the orbital direction of the distribution function) is the lattice speed of the D2Q9 model. $\boldsymbol{M}$ is an orthogonal transformation matrix, and $\mathbf{\Lambda}$ is a diagonal matrix.[19] $\boldsymbol{F}$ denotes the body force on the fluid, including the inter-particle interaction force $\boldsymbol{F}_{int}$, the adhesion force between liquid/vapor and solid $\boldsymbol{F}_{ads}$, and the gravitational force $\boldsymbol{F}_g$,

$$\boldsymbol{F}_{int} = -G\psi(\boldsymbol{x})\sum_{\alpha=1}^8 w_\alpha \psi(\boldsymbol{x} + \boldsymbol{e}_\alpha \Delta t)\boldsymbol{e}_\alpha,$$
$$\boldsymbol{F}_{ads} = -G_w \psi(\boldsymbol{x})\sum_{\alpha=1}^8 w_\alpha s(\boldsymbol{x} + \boldsymbol{e}_\alpha \Delta t)\boldsymbol{e}_\alpha, \tag{2}$$
$$\boldsymbol{F}_g = (\rho - \rho_{ave})\boldsymbol{g},$$

where $w_\alpha$ are the weighting coefficients ($w_0 = 4/3$, $w_{1-4} = 1/3$, and $w_{5-8} = 1/12$). $\rho_{ave}$ is the average density of the whole fluid domain, and $\boldsymbol{g} = (0, -g)$ is the gravitational acceleration. The desired contact angle can be obtained by changing the fluid–solid interaction strength $G_s$. $s(\boldsymbol{x})$ is the indicator function equal to 1 when $\boldsymbol{x}$ is a solid node and 0 when $\boldsymbol{x}$ is a fluid node,

$$\psi(\boldsymbol{x}) = \sqrt{\frac{2(p_{EOS} - \rho c_s^2)}{G}},$$
$$p_{EOS} = \frac{\rho RT}{1 - b\rho} - \frac{a\rho^2 \alpha(T)}{1 + 2b\rho - b^2\rho^2}, \tag{3}$$
$$\alpha(T) = [1 + (0.37464 + 1.54226\omega - 0.26992\omega^2)$$
$$\times (1 - \sqrt{T/T_c})]^2,$$

where $p_{EOS}$ represents Peng–Robinson (P–R) equation of state. $\omega$ is the central factor, and $R$ is the gas constant. This study sets $a = 3/49$, $b = 2/21$, $R = 1$, and $\omega = 0.344$, corresponding to the critical temperature $T_c = 0.1094$. More details about the lattice Boltzmann method can be found in a previous study.[20–23]

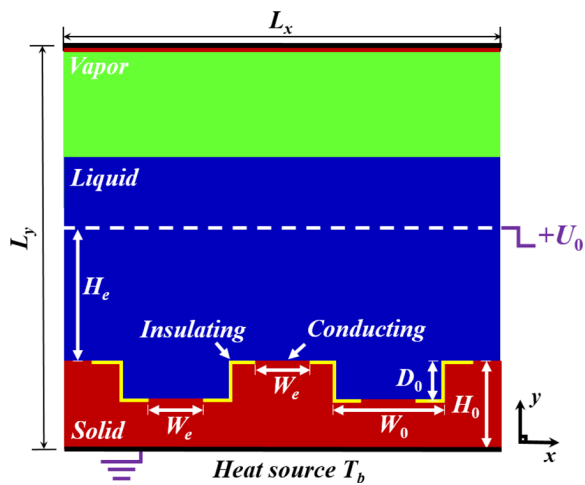

**FIG. 1.** Schematic of the conducting–insulating microcavity surfaces under the action of an electric field.

## C. Energy equation

According to the local balance law for entropy, the energy equation considering the vapor–liquid phase change can be expressed as follows:[23]

$$\frac{\partial T}{\partial t} = -\boldsymbol{u}\nabla T + \frac{1}{\rho c_v}\nabla \cdot (\lambda \nabla T) - \frac{T}{\rho c_v}\left(\frac{\partial p_{EOS}}{\partial T}\right)_\rho \nabla \cdot \boldsymbol{u}, \quad (4)$$

where $T$, $c_v$, and $\lambda$ denote the temperature, heat capacity, and thermal conductivity, respectively. Computing Eq. (4) by the lattice Boltzmann method is the typical strategy.[24,25] However, it is still necessary to introduce the finite difference method to compute the gradient terms involved in this process. To simplify the numerical model, the finite difference method is directly used to calculate the energy equation in this study. The right-hand side of Eq. (4) can be expressed in terms of $K(T)$. The discretization of the time term is carried out using a fourth-order Runge–Kutta scheme,

$$T^{t+\Delta t} = T^t + \frac{\Delta t}{6}(h_1 + h_2 + h_3 + h_4), \quad (5)$$

where $h_1$, $h_2$, $h_3$, and $h_4$ are given by the following equations:

$$h_1 = K(T^t), \quad h_2 = K\left(T^t + \frac{\Delta t}{2}h_1\right),$$
$$h_3 = K\left(T^t + \frac{\Delta t}{2}h_2\right), \quad h_4 = K(T^t + \Delta t h_3), \quad (6)$$

the gradient in Eq. (4) is calculated using an isotropic difference format,

$$\partial_i \phi(\boldsymbol{x}) \approx \frac{1}{c_s^2}\sum_\alpha w_\alpha \phi(\boldsymbol{x} + \boldsymbol{e}_\alpha \Delta t)\boldsymbol{e}_\alpha. \quad (7)$$

In previous numerical simulations, the conjugate heat transfer at the solid–liquid interface is usually neglected to simplify the calculations and the constant temperature boundary conditions are directly used, which leads to missing information about the thermal response of the solid and the heat flux. However, the problem of energy non-conservation or numerical divergence occurs when the diffusion term $\nabla\cdot(\lambda\nabla T)$ is solved directly by the finite difference method. In this study, the numerical format suggested by Hu and Liu et al.[26] is used to compute the diffusion term. The control volume $\delta_v$ for any point $P(x, y)$ in the computational domain has the following equation:

$$\nabla \cdot (\lambda \nabla T)_P = \frac{q_{\text{diff},P}}{\delta_v} = \frac{\displaystyle\int_{\delta_v} \nabla \cdot (\lambda \nabla T) dV}{\Delta x \Delta y},$$

$$\int_{\delta_v} \nabla \cdot (\lambda \nabla T) dV = \Delta x \Delta y \left[\lambda_e\left(\frac{\partial T}{\partial x}\right)_e - \lambda_w\left(\frac{\partial T}{\partial x}\right)_w \right. \quad (8)$$
$$\left. + \lambda_n\left(\frac{\partial T}{\partial y}\right)_n - \lambda_s\left(\frac{\partial T}{\partial y}\right)_s\right],$$

where the lattice step $\Delta x = \Delta y = 1$. To ensure energy conservation, the average thermal conductivity can be expressed as

$$\lambda_e = \frac{2\Delta x}{\dfrac{\Delta x}{\lambda_x} + \dfrac{\Delta x}{\lambda_{x+1}}}, \quad \lambda_w = \frac{2\Delta x}{\dfrac{\Delta x}{\lambda_x} + \dfrac{\Delta x}{\lambda_{x-1}}},$$
$$\lambda_n = \frac{2\Delta y}{\dfrac{\Delta y}{\lambda_y} + \dfrac{\Delta y}{\lambda_{y+1}}}, \quad \lambda_s = \frac{2\Delta y}{\dfrac{\Delta y}{\lambda_y} + \dfrac{\Delta y}{\lambda_{y-1}}}. \quad (9)$$

A first-order difference format is used to obtain the temperature gradient at the $P$-point in different directions,

$$\left(\frac{\partial T}{\partial x}\right)_e = \frac{T_{x+1} - T_x}{\Delta x}, \quad \left(\frac{\partial T}{\partial x}\right)_w = \frac{T_x - T_{x-1}}{\Delta x},$$
$$\left(\frac{\partial T}{\partial y}\right)_n = \frac{T_{y+1} - T_y}{\Delta y}, \quad \left(\frac{\partial T}{\partial y}\right)_s = \frac{T_y - T_{y-1}}{\Delta y}. \quad (10)$$

## D. Electric field model

In classical electromagnetic theory, the electrical force on an incompressible fluid is given by the following equation:[27–29]

$$\boldsymbol{F}_e = \rho_e \boldsymbol{E} - \frac{1}{2}E^2\nabla\varepsilon, \quad (11)$$

where $\rho_e$, $\boldsymbol{E}$, and $\varepsilon$ denote the free charge density, electric field intensity, and dielectric constant, respectively. The first and second terms on the right-hand side of the equation are the Coulomb and polarization forces, respectively. For FC-72, the charge relaxation time $\tau(\sim100\,\text{s})$ is much larger than the bubble growth time $t(\sim0.01\,\text{s})$, which implies that there is almost no free charge at the interface during bubble motion.[16] Therefore, the Coulomb force can be neglected. In this study, only the effect of polarization force on boiling heat transfer is considered,

$$\boldsymbol{F}_e = -\frac{1}{2}E^2\nabla\varepsilon, \quad (12)$$

the governing equation of the electric field neglecting charge motion can be expressed as

$$\nabla \cdot (\varepsilon \boldsymbol{E}) = 0, \quad (13)$$

where $\varepsilon$ denotes the dielectric constant, and $\boldsymbol{E}$ is the electric field intensity. When the electric field vectors are irrotational, the electric field intensity $\boldsymbol{E}$ is related to the electric potential gradient (i.e., $\boldsymbol{E} = -\nabla U$). Equation (13) can be rewritten as follows:

$$\nabla \cdot (\varepsilon \nabla U) = 0. \quad (14)$$

The electric potential distribution in the system can be obtained by solving Eq. (14) by the finite difference method.

## E. Simulation setup and boundary conditions

The characteristic length $l_0$ and time $t_0$ are typical dimensions to describe the boiling process,

$$l_0 = \sqrt{\frac{\sigma}{(\rho_l - \rho_v)g}}, \quad t_0 = \sqrt{l_0/g}. \quad (15)$$

The boiling heat transfer performance can be assessed from the dimensionless heat flux $Q''(x,t)$ at the bottom of the heater and the surface superheat $Ja$,

$$Q''(x,t) = \frac{\int_{t_a}^{t_b} Q'(t)dt}{t_b - t_a}, \quad Ja = \frac{c_p(T_w - T_{sat})}{h_{lv}}, \qquad (16)$$

where $t_a$ to $t_b$ are time intervals from 20 000 to 50 000 t.s. (t.s. denotes a time step). $T_w$ is the space- and time-averaged wall temperature over the heating surface. $Q'(x,t)$ is the space-averaged dimensionless heat flux,

$$Q'(x,t) = -\frac{\lambda_s \cdot l_0}{\mu_l h_{lv}} \cdot \frac{\int_0^{L_x} \nabla_y T|_{x,0} dx}{L_x}. \qquad (17)$$

The temperature gradient $(\nabla_y T|_{x,0})$ at the bottom of the heater can be obtained in a second-order differential format.

The simulation grid size was $L_x \times L_y = 400$ l.u. $\times$ 600 l.u. (l.u. means lattice units), and this computational domain ensures the grid independence. The initial fluid domain with $y < 0.70L_y$ is saturated liquid, and $y > 0.70L_y$ is saturated vapor. The saturation temperature is $T_{sat} = 0.867_c$. The equilibrium densities of the liquid and vapor were obtained by Maxwell's construction as $\rho_l = 6.5$ and $\rho_v = 0.38$, respectively. The structural parameters of the conducting–insulating microcavity surface were as follows: $H_0 = 50$ l.u., $W_0 = 100$ l.u., $D_0 = 10$ l.u., and $W_e = 50$ l.u. The thermal conductivity ratio between solid and liquid is $\lambda_s/\lambda_l = 50$. The specific heat at a constant volume is $c_{V,s} = c_{V,l} = c_{V,v} = 6.0$. The solid density is $\rho_s = 1.5\rho_l$. The dynamic viscosity ratio and thermal conductivity ratio of the liquid to the vapor are $\mu_l/\mu_v = 17$ and $\lambda_l/\lambda_v = 17$, respectively. The static contact angle is $\theta = 30.0°$, and the gravitational acceleration is $\mathbf{g} = (0, -0.000 03)$. The latent heat of the vapor–liquid phase change was determined to be $h_{lv} = 0.58$ from the theoretical analysis.[30] The vapor–liquid surface tension coefficient obtained via Laplace's law test was $\gamma = 0.086$. According to Eq. (15), the characteristic length $l_0$ is 21.64 l.u. and the time $t_0$ is 849 t.s. If not specified, the electric field intensity $E_0 = 0.08$ remains constant. Referring to the study by Feng et al.,[31] the dielectric constants of vapor and liquid are $\varepsilon_v = 1.0$ and $\varepsilon_l = 2.24$, respectively. The parameters used in this study are all in lattice units.

Periodic conditions (including flow, temperature, and electric fields) are imposed on the left and right boundaries of the computational domain. The halfway bounce-back format is imposed on the density distribution function to realize the no-slip condition at the solid–liquid interface. Constant temperature and constant electric potential boundaries are used at the bottom and top of the computational domain,

$$T(x, y = 0, t) = T_b, \quad U(x, y = \Omega_s, t) = 0,$$
$$T(x, y = L_y, t) = T_{sat}, \quad U(x, y \geq H_e, t) = U_0, \qquad (18)$$

where $\Omega_s$ denotes the $y$-axis coordinates of the microcavity surface. The insulating can be achieved by setting the electric potential gradient to zero for the corner wall. Therefore, the electric potential at the insulated wall can be obtained by the following equation:

$$U(\mathbf{x}) = \frac{\sum_{\alpha}^{8} w_{\alpha}(1 - s(\mathbf{x} + \mathbf{e}_{\alpha}\Delta t))U(\mathbf{x} + \mathbf{e}_{\alpha}\Delta t)}{\sum_{\alpha}^{8} w_{\alpha}(1 - s(\mathbf{x} + \mathbf{e}_{\alpha}\Delta t))}. \qquad (19)$$

## III. MODEL VALIDATION

### A. Contact angle test

The fluid–solid interaction strength $G_w$ determines the magnitude of the force between the solid wall and the liquid, which will change the contact angle of the droplet. Figure 2 quantitatively gives the relationship between contact angle and $G_w$. Initially, a circular droplet of radius $R_d = 15$ l.u. is in contact with the wall. The results of the numerical calculations after reaching stabilization are shown in the inset. The simulation neglects the effect of gravity, so the contact angle is determined only by the adhesion force between liquid and solid $F_{ads}$. When the strength $G_w$ is greater than zero, the solid and liquid repel each other and the wall is hydrophobic. Conversely, the solid and the liquid attract each other and the wall is hydrophilic. When the strength $G_w$ is equal to zero, the wall is neutral, i.e., the contact angle is 90°. This phenomenon is consistent with previous results.[32]

### B. Validation of the boiling model

The boiling curve on a smooth surface is simulated and compared with the classical correlation equation to verify the correctness of the numerical model of phase change heat transfer developed in this study. As the superheat $Ja$ increases, the heating surface sequentially experiences natural convection, the onset of nucleate boiling (ONB), and the CHF, as shown in Fig. 3. The boiling curve of the smooth surface obtained from lattice Boltzmann simulations is compared with the Rohsenow's correlation equation,[33]

$$Ja = C_{sf}\left(\frac{Q}{\mu_l h_{lv}}\sqrt{\frac{\sigma}{g(\rho_l - \rho_v)}}\right)^{0.33}\left(\frac{C_{p,l}\mu_l}{\lambda_l}\right)^{1.7}, \qquad (20)$$

where $C_{sf}$ is a fitting coefficient in the range of 0.006–0.0130 and related to the physical properties of the heating surface. The red solid line in Fig. 2 indicates the fitting result with $C_{sf} = 0.0068$. The heat flux obtained from the simulation is consistent with the Rohsenow's correlation equation. Zuber proposed an empirical correlation formula for CHF,[34]

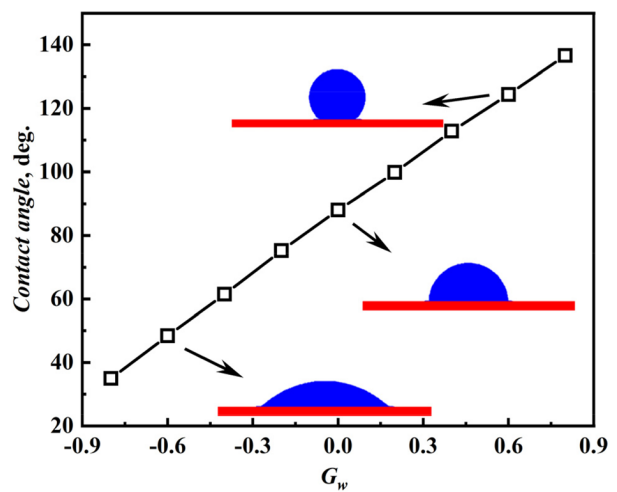

**FIG. 2.** Contact angle vs fluid–solid interaction strength $G_w$.

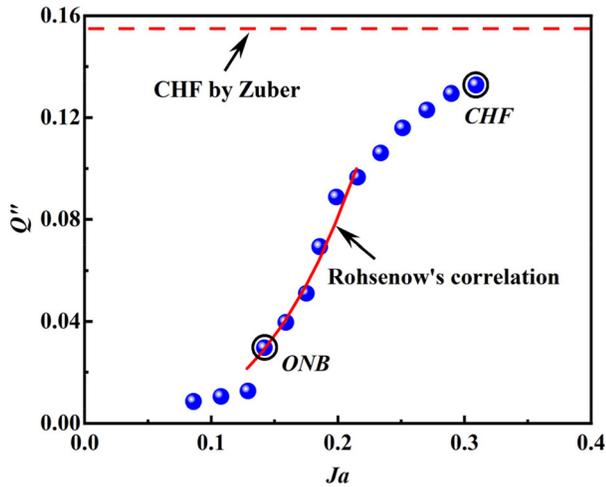

**FIG. 3.** Boiling curves for smooth surfaces.

$$Q_{CHF} = K\rho_v h_{fg} \left[\frac{\sigma(\rho_l - \rho_v)g}{\rho_v^2}\right]^{-0.5}, \qquad (21)$$

where $K$ is the fitting coefficient between 0.12 and 0.16.[35] $Q_{CHF}$ can be further expressed in the dimensionless form $Q_{CHF}^r = Q_{CHF} \cdot l_0/\mu_l h_l$, $Q_{CHF}^r = 0.133$ obtained from the numerical simulation is closer to the predicted value 0.155 ($K = 0.12$), which verifies the correctness of the model.

### C. Deformation of droplets in the electric field

The deformation of a static droplet under the action of an electric field is simulated to verify the correctness of the electric field model. A circular droplet of radius $R_d = 25$l.u. is placed statically in a square cavity with a computational domain of $4R_d \times 4R_d$. A high electric potential is maintained at the top of the square cavity, and the bottom of the square cavity is grounded. The deformation of the droplet under the electric field is defined as $D = (L-B)/(L+B)$, where $L$ and $B$ are the lengths of the long and short axes, respectively. Allan and Mason[36] provided an analytical solution for the droplet deformation $D^*$,

$$D^* = \frac{9}{16} Ca_E \frac{(\varepsilon_r - 1)^2}{(\varepsilon_r + 2)^2}, \qquad (22)$$

where $Ca_E (= \varepsilon_v R_d E^2/\gamma)$ is the electrical capillary number, and $\varepsilon_r (= \varepsilon_l/\varepsilon_v)$ is the dielectric constant ratio.

Table I gives the results of the numerical solution compared to the analytical solution. The droplet deformation increases as the dielectric constant ratio and the electric capillary number increase. The numerical results obtained in this study are close to the theoretical values when the dielectric constant ratio is in the range of 1.5–2.5.

### IV. RESULTS AND DISCUSSION

This section gives the boiling curves for the microcavity surface (MC), the uniformly conducting microcavity surface (MC-UCS), and the conducting–insulating microcavity surface (MC-CIS). The heat transfer mechanism of the heating surface under the action of electric field (EF) was revealed by analyzing the bubble behavior and the thermal response of the heater bottom. Finally, the effects of different electric field intensities on CHF were explored, and a new predictive correlation equation was established.

**TABLE I.** Effect of electrical capillary number on droplet deformation.

| $\varepsilon_l$ | $\varepsilon_v$ | $Ca_E$ | $D$ | $D^*$ |
|---|---|---|---|---|
| 1.5 | 1 | 0.3 | 0.003 92 | 0.003 44 |
| 1.5 | 1 | 0.6 | 0.007 14 | 0.006 89 |
| 2.0 | 1 | 0.3 | 0.011 00 | 0.010 55 |
| 2.0 | 1 | 0.6 | 0.022 13 | 0.021 09 |
| 2.5 | 1 | 0.3 | 0.020 36 | 0.018 75 |
| 2.5 | 1 | 0.6 | 0.039 25 | 0.037 50 |

### A. Boiling curves

The boiling curves for the three heating surfaces (MC, MC-UCS, and MC-CIS) are shown in Fig. 4. The beginning and end of each boiling curve are ONB and CHF, respectively. As can be seen in the figure, there are two different regimes for the electric field effect on the boiling curve. In the low superheat boiling regime, the heat flux of MC-UCS is lower than that of MC due to the field trap effect, which is consistent with the experimental results of Quan et al.[15] Specifically, the heat flux of the MC after electric field activation at $Ja = 0.15$ decreased from 0.068 to 0.053, which is a 28.3% reduction. At the same superheat, the MC-CIS proposed in this study reduces the heat flux on the microcavity surface under the electric field to only 0.062, which is a reduction of 9.7%. This phenomenon indicates that the inhibition of heat transfer from the microcavity surface by the electric field is somewhat improved in the boiling regime with low superheat.

By further increasing the superheat ($Ja > 0.20$), the microcavity surface enters a fully developed boiling regime. The boiling curves of both MC-UCS and MC-CIS are shifted to the upper left side of MC, which indicates that the electric field enhances the boiling heat transfer performance of the microcavity surface at high superheat. However, this trend of heat transfer enhancement varies across different heating surfaces. For MU-UCS, the boiling curve gradually flattens out with increasing superheat, while the slope of the boiling curve for MU-CIS is much larger and remains almost constant. As a result, MU-CIS can maintain relatively high boiling heat transfer coefficients even when the heating surface is under high superheat.

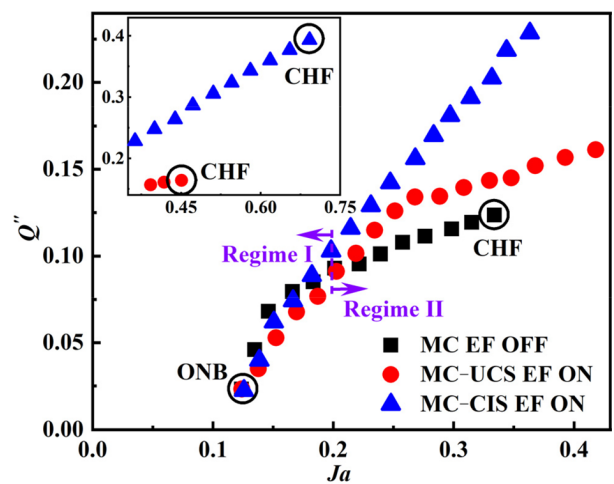

**FIG. 4.** Boiling curves for different surfaces.

The CHFs of the three boiling surfaces are given in Fig. 5. The CHFs of MC, MC-UCS, and MC-CIS are 0.123, 0.164, and 0.393, respectively, which suggests that the introduction of an electric field on the surface of the microcavity can enhance the CHF. However, the presence of the field trap effect for a given electric field intensity resulted in only a 33.3% improvement of MC-UCS over MC. For MC-CIS, when the electric field distribution was reorganized, the CHF increased by 219.5% compared to MC. In this study, we will further investigate the thermodynamic mechanism of the effect of electric field on the boiling heat transfer performance of microcavity surfaces under different superheat.

## B. Bubble behaviors

The iso-potential lines and electric field intensity distributions of MC-UCS and MC-CIS are given in Fig. 6 (only the electric field distribution was calculated numerically) to understand the mechanism of synergistic enhancement of boiling heat transfer by the electric field and microcavity surfaces. As can be seen in the figure, the microcavity surface changes the form of the distribution of the electric field in space. For MC-UCS, the heating surface maintains a constant electric potential, which causes significant distortion of the iso-potential lines at the corners of the top of the microcavity, which locally generates a large electric field intensity (i.e., $E = -\nabla U$). The iso-potential lines are almost horizontally distributed at locations away from the microcavity surface. The phenomenon is similar to the numerical results of Liu *et al.*[16] and Li *et al.*[17]

The reorganization of the spatial electric field will affect the characteristics of the electrical force distribution at the vapor–liquid interface, which will change the bubble dynamics. According to Eq. (12), the electrical force shows a positive correlation with the electric field intensity, which means that the bubbles in the cavity may be continuously suppressed on the heating surface by the strong electrical force induced by the high electric field intensity at the corners (i.e., the field trap effect). For the MC-CIS proposed in this study, since the corners of the microcavity are electrically insulated, the electric potential gradient in the direction perpendicular to the wall is zero, which produces a low electric field intensity locally. Moreover, iso-potential line changes dramatically at the junction of the conducting and insulating surfaces, resulting in a significant increase in the electric field intensity compared with the MC-UCS. This phenomenon leads to the elimination of the field trap effect on the microcavity surface. The differences in the boiling characteristics of the microcavity surface with and without the field trap effect will be discussed further.

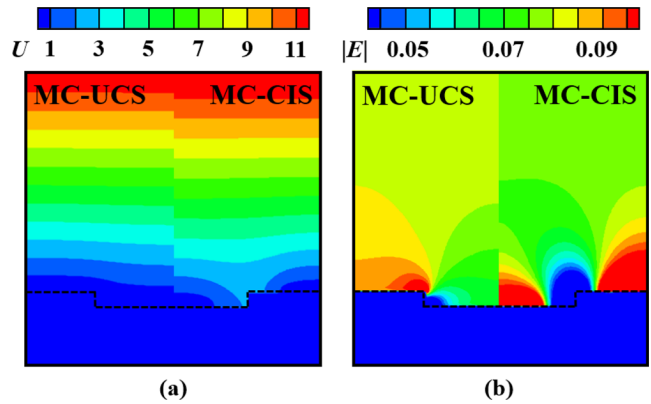

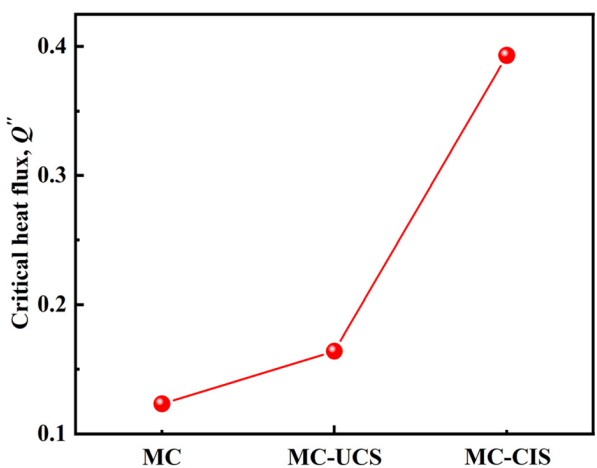

**FIG. 5.** CHF on different surfaces.

**FIG. 6.** Distribution of iso-potential line $U$ and electric field intensity $|E|$: (a) iso-potential line and (b) electric field intensity. The black dashed line indicates the solid–liquid interface.

Figure 7 shows the boiling behavior of the three surfaces for $T_b = 0.102T_c$ ($Ja \approx 0.15$). (Time intervals are located within the first three boiling cycles of the heating surfaces. A complete boiling cycle involves the nucleation, growth, and departure of bubbles.) The bubble is the blue region, and the solid heater is the red region. As shown in Fig. 7(a), due to the fact that the top of the microcavity is farther away from the heat source, this results in fewer bubble nucleation sites than that in the interior of the microcavity. Specifically, there are three nucleation sites inside the microcavity at $t^* = 35.3$, while there is only one at the top. In addition, the corner structure of the microcavity accelerated the nucleation of the bubbles so that they detached from the heating surface earlier. Subsequently, the residual vapor left by bubble detachment at the corners of the microcavity continued to grow, producing a lateral merger between the bubbles at $t^* = 47.1$. The large bubbles produced by the merger at $t^* = 58.9$ were detached from the heating surface by buoyancy.

As shown in Fig. 7(b), for MC-UCS, the high electric field intensity present at the top surface of the microcavity accelerates bubble departure, but bubble growth inside the microcavity is inhibited by the field trap effect. At $t^* = 35.3$, the bubble necks at the corners are just forming, and the bubble sizes in the center region of the microcavity are small. However, for the case where the electric field is not activated, the first boiling cycle in the microcavity has been completed (i.e., the bubble neck is broken) and the size of the bubbles in the center region of the microcavity is larger, as shown in Fig. 7(a). Prolonged attachment of bubbles to the bottom of the microcavity causes localized hot spots on the heating surface, negatively affecting the boiling heat transfer performance. Thus, while heat transfer at the top of the cavity under the action of the electric field is improved, the greater number of bubbles in the cavity allows the field trap effect to dominate the boiling performance of the overall heating surface. This phenomenon results in the boiling curve of MC-UCS being on the right side of MC at low superheat boiling regime ($0.13 < Ja < 0.20$), as shown in Fig. 4.

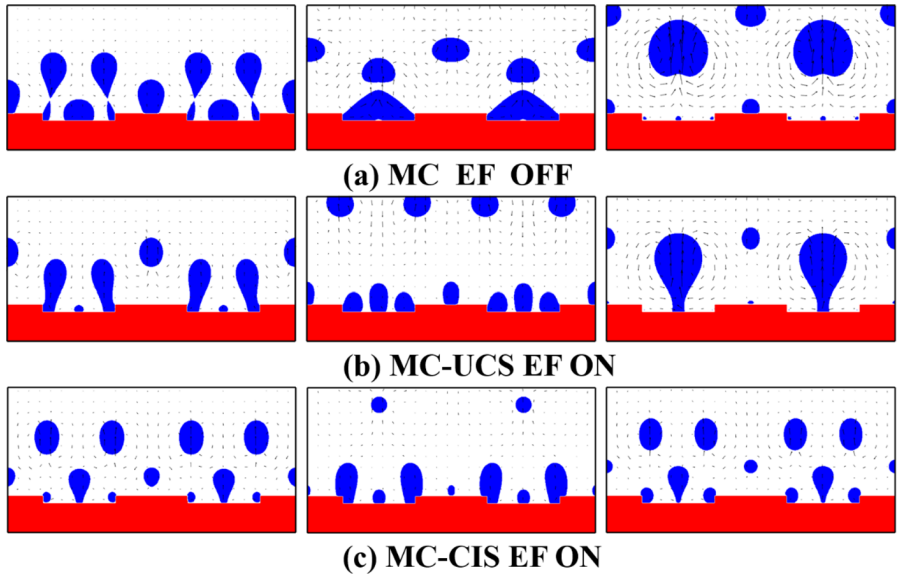

As shown in Fig. 7(c), for MC-CIS, the bubbles in the center region of the microcavity are about to detach from the heating surface at $t^* = 35.3$ driven by the electrical force at the bottom, and the bubbles at the corners have already detached. Therefore, when the field trap effect is eliminated, the bubble generation period is reduced compared to MC-UCS, which will enhance the boiling performance to some extent. However, the numerical results show that the boiling performance of MC-CIS is still slightly weaker than that of MC (Fig. 4). For MC, small bubbles tend to merge into large bubbles, which cause strong convection when they depart from the heating surface, which facilitates the rapid cooling of the heating surface by the low-temperature bulk liquid [Fig. 7(a), $t^* = 58.9$]. For MC-CIS, the electrical force in the microcavity hinders the lateral movement of the bubbles, and the bubbles do not merge with each other, making the convection effect weak [Fig. 7(c), $t^* = 58.9$]. In summary, the boiling heat transfer performance is MC > MC-CIS > MC-UCS at low superheat.

As the superheat increases, the heating surface enters a fully developed boiling regime. Figure 8 shows the boiling behavior of different surfaces at $T_b = 0.120\,T_c$ ($Ja \approx 0.30$). For MC, the cavity is filled with vapor when the superheat is larger. The three-phase contact line is always pinned at the cavity mouth during the complete boiling cycle, as shown in Fig. 8(a). Vapors trapped in the cavity increase the thermal resistance of the heating surfaces and impede the return of the bulk liquid. As a result, in the fully developed boiling regime, the boiling curve of MC exhibits a more flat growth trend, as shown in Fig. 4.

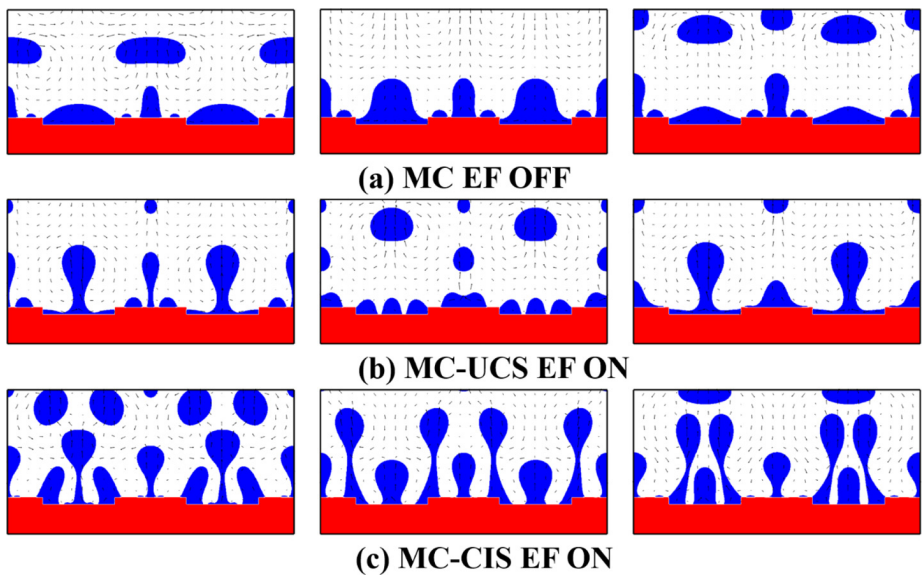

Experimental studies[15] showed that electric fields can improve the heat transfer performance of heating surfaces by destabilizing the vapor–liquid interface. A similar phenomenon was observed in numerical simulations. For MC, the vapor block inside the microcavity was always stable over the heating surface during the complete boiling cycle [Fig. 8(a)]. For MC-UCS, the interface breaks up into three isolated small bubbles at the position of the thinnest vapor block when the large bubbles detach [Fig. 8(b), $t^* = 35.3$–47.1]. The electrical force and electric field intensity distributions of the MC-UCS at $t^* = 35.3$ are given in Fig. 9(a) to explain the breakup mechanism of the vapor block. Since the dielectric constant of the liquid is greater than that of the vapor, there is an electrical force at the interface pointing toward the interior of the bubble. (The dielectric constant gradient $\nabla \varepsilon$ determines the direction of the electrical force.) Equation (14) shows that the magnitude of the electric potential is related to the dielectric constant, which leads to the vapor–liquid interface morphology to change the spatial electric field distribution. The iso-potential line changes dramatically at the location of the thinnest vapor block within the microcavity, which locally generates a larger electric field intensity and increases the squeezing of the vapor block by the electrical force, thus promoting interfacial breakup, as shown in Fig. 9(a). Therefore, MC-UCS at high superheat may improve boiling performance by promoting bubble breakup near the heating surface.

No vapor block appears in the microcavity at $T_b = 0.120T_c$ for the MC-CIS, as shown in Fig. 8(c). At the junction of the conducting and insulating surfaces, the electrical force increased significantly [Fig. 9(b)], which caused the pinning of the contact line, thus hindering the merging of the bubbles in the cavity. This phenomenon provides multiple continuously open vapor–liquid separation paths on the heating surface, which effectively enhances the convective heat transfer in the cavity and delays the occurrence of CHF.

As the superheat increases, the MC is covered by a complete vapor film, thus reaching the CHF. The vapor–liquid interfacial morphology (white dashed line) and temperature field of MC-UCS and MC-CIS at $T_b = 0.134T_c$ are given in Fig. 10. For MC-UCS, the larger vapor pressure inside the microcavity at sufficiently high superheat offsets the electrical force, allowing the perturbation of the interface to be suppressed. As a result, the vapor block stably covered the microcavity without breakup, as shown in Fig. 10(a). This leads to a significant increase in the thermal resistance of the heating surface, which is not conducive to boiling heat transfer. On the other hand, the strong electrical force at the corner of the microcavity avoids the interconnection of vapor blocks into a vapor film by squeezing the vapor–liquid interface, thereby improving CHF (Fig. 4).[17]

For MC-CIS, the contact lines are still pinned at the junction of the conducting and insulating surfaces, as shown in Fig. 10(b). From the figure, it can be seen that under the squeezing of electrical force and the lifting of buoyancy, the bubbles appear rod-shaped and detach away from the heating surface. Figure 11 shows the heater bottom time-averaged temperatures of the MC-UCS and MC-CIS at $T_b = 0.134T_c$ ($t^* = 35.3$–58.9). At the bottom of the microcavity (50 l.u.$< x <$150 l.u. and 250 l.u.$< x <$350 l.u.), the highest and lowest temperatures were observed for MC-UCS and MC-CIS, respectively, indicating that the proposed novel heating surface improves the heat transfer in the cavity. At the top surface of the microcavity (150 l.u.$< x <$250 l.u.), the non-uniform electrical force generated by the MC-CIS promotes bubble departure and facilitates the re-wetting of the wall by the bulk liquid, as shown in Fig. 10(b). As a result, a localized low-temperature region appears on the top surface of the microcavity (Fig. 11). For MC-UCS, the electrical force tends to act uniformly on the vapor–liquid interface, which is detrimental to the bubble departure and increases the localized temperature. In conclusion, the boiling heat transfer performance is MC-CIS > MC-UCS >MC at high superheat (Fig. 4).

### C. Effect of electric field intensity on boiling curves with MC-CIS

Based on the above results, it can be seen that MC-CIS has better boiling performance compared to MC and MC-UCS. The effect of different electric field intensities on the boiling curves of MC-CIS is further discussed in this section, as shown in Fig. 12. Gong et al.[37] found a second transition point in the boiling curve (the slope of the boiling curve starts to decrease) as the superheat increases, which comes from the formation of permanent dry spots. The figure shows that different electric field intensities have less effect on the boiling curve before the appearance of the second transition point. However, a larger electric field intensity can delay the superheat needed for the second transition point to occur. The superheat required to form the second transition point is 0.18, 0.28, and 0.35 at $E_0 = 0.0$, 0.050, and 0.065, respectively. Particularly, a sufficiently large electric field intensity ($E_0 = 0.080$) can cause the second transition point to disappear. This is because the electrical force promotes the contraction of the interface, thereby hindering the continuous adhesion of large dry spots on the heating surface.

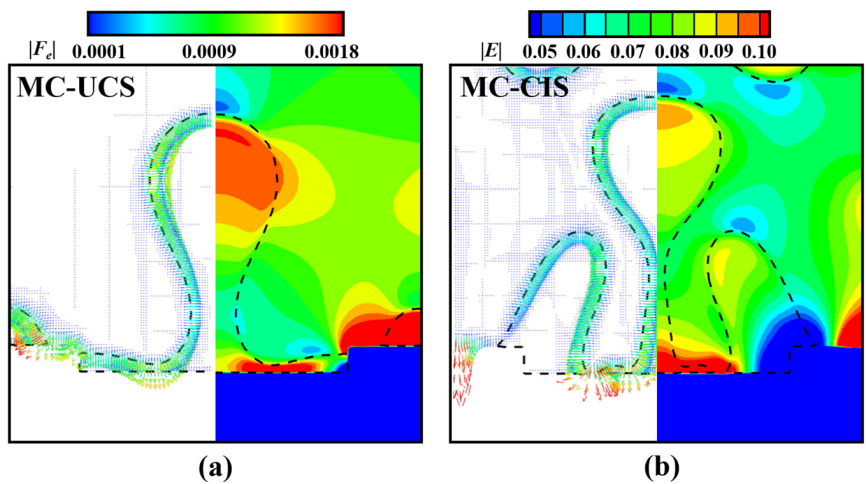

**FIG. 9.** Distributions of electrical force $|F_e|$ and electrical field intensity $|E|$ during bubble departure ($T_b = 0.120T_c$ and $t^* = 35.3$). (a) The uniformly conducting microcavity surface after the electric field is turned on. (b) The conducting–insulating microcavity surface after the electric field is turned on.

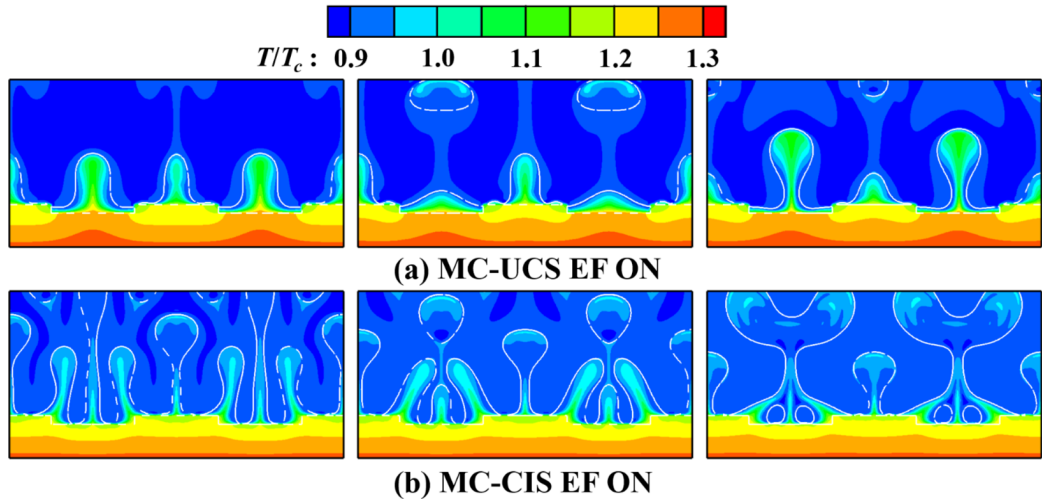

**FIG. 10.** Temperature fields in the computational domain at $T_b = 0.134\ T_c$ (from left to right: $t^* = 35.3$, 47.1, and 58.9). (a) The uniformly conducting microcavity surface after the electric field is turned on. (b) The conducting–insulating microcavity surface after the electric field is turned on.

CHF is the main factor limiting the improvement of the cooling performance of electronic chips, so the prediction of CHF is especially important. The CHF correlation for a smooth surface in a uniform electric field was proposed by Johnson,[38]

$$r_{CHF} = Q''_{CHF,E}/Q''_{CHF,0} = \sqrt{\frac{El^* + \sqrt{El^* + 3}}{\sqrt{3}}},$$

$$El^* = \frac{\varepsilon_l(\varepsilon_l - \varepsilon_v)E_0^2}{\varepsilon_v(\varepsilon_l + \varepsilon_v)\sqrt{(\rho_l - \rho_v)\sigma g}},$$

(23)

where $r_{CHF}$ is the enhancement ratio of CHF under a uniform electric field, and $Q''_{CHF,0}$ is the CHF at $E_0 = 0$. For conducting–insulating microcavity surfaces, the iso-potential lines are reorganized, which changes the spatial electric field intensity distribution. Therefore, Eq. (23) needs to be corrected to account for structural effects on the electric field. Based on the analysis in Sec. IV B, it can be seen that a lower electric field intensity exists at the corners, but it does not negatively affect the boiling heat transfer. On the other hand, an electric field intensity much higher than that of the smooth surface is generated at the junction of the conducting and insulating surfaces. Therefore, the effective electric field intensity $E_e$ can be regarded as enhanced compared to smooth surface. The effective electric field intensity $E_e$ can be decomposed into two parts. The first part is the electric field intensity $E_0$ on a smooth surface. The second part is the additional electric field intensity $\psi E_0$ caused by local iso-potential line changes. The corrected electric field intensity is as follows:

$$E_e = E_0(1 + \psi), \tag{24}$$

where $\psi$ is the fitting coefficient related to the structure of the heating surface. Therefore, the modified $El^*$ can be expressed as follows:

$$El^* = \frac{(1+\psi)^2 \varepsilon_l (\varepsilon_l - \varepsilon_v) E_0^2}{\varepsilon_v (\varepsilon_l + \varepsilon_v) \sqrt{(\rho_l - \rho_v)\sigma g}}. \tag{25}$$

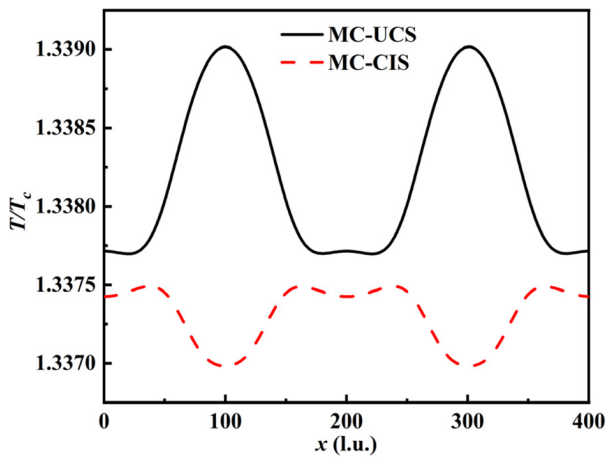

**FIG. 11.** Time-averaged temperatures at the bottom of the heaters of MC-UCS and MC-CIS at $T_b = 0.134\ T_c$ ($t^* = 35.3$–58.9).

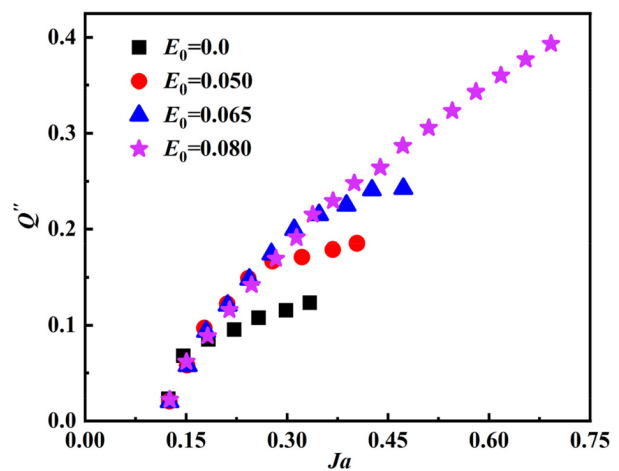

**FIG. 12.** Boiling curves of MC-CIS at different electric field intensities.

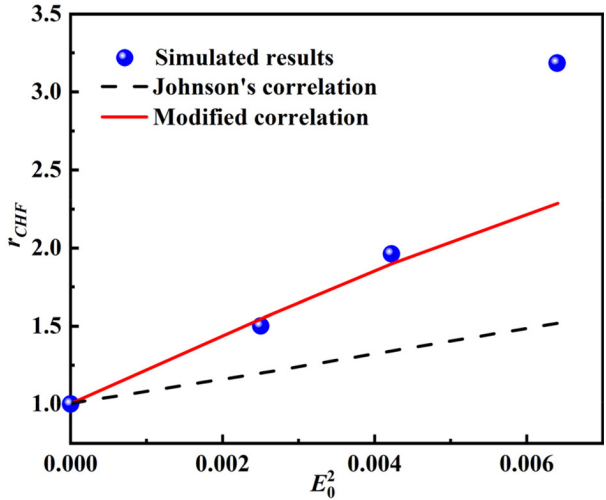

FIG. 13. Effect of electric field intensity on the enhancement ratio $r_{CHF}$.

Figure 13 gives the effect of the square of the electric field intensity $E_0^2$ on the enhancement ratio $r_{CHF}$ and compares the Johnson correlation and modified results. When $E_0^2 < 0.005$, $r_{CHF}$ increases linearly with $E_0^2$. Numerical results are gradually larger than predicted values when $E_0^2 > 0.005$. Given that the Johnson correlation equation does not consider the effect of microstructure on the electric field enhancement, predicted values become lower. For the modified correlation equation proposed in this study, the predicted values are in good agreement with the numerical results at $E_0^2 < 0.005$ ($\psi$ taken as 0.64).

This study further analyzes why the numerical results are larger than the predicted values at high electric field intensity. The CHF correlation [i.e., Eq. (23)] given by Johnson[38] was obtained based on the assumption that the electrical force acts uniformly at the vapor–liquid interface (i.e., the electric field reduces only the Taylor wavelength). The above-mentioned assumptions are reasonable for uniform electric fields on smooth surfaces, but some limitations are identified for non-uniform electric fields on surfaces with microstructures. Figure 14 demonstrates the interface evolution of the heating surface in the CHF state at different electric field intensities. When the electric field intensity $E_0 = 0$, the microcavity is covered by a stable vapor block. As the electric field intensity increases ($0 < E_0 \leq 0.065$), the vapor block in the microcavity shows intermittent breakup and regeneration under the perturbation of the electrical force, which indicates that the electric field enhances the CHF by decreasing the Taylor wavelength. When the electric field intensity is sufficiently large ($E_0 = 0.080$), pinning is produced at the junction of the conducting and insulating surfaces. The development of capillary waves in the microcavity was suppressed because of the non-uniform distribution of the electrical force. This phenomenon leads to the reflux of the bulk liquid and the formation of vapor jets to become well-ordered, significantly increasing the CHF. Thus, the modified correlation equation underestimates the enhancement ratio $r_{CHF}$ when the electric field intensity is sufficiently large (Fig. 13).

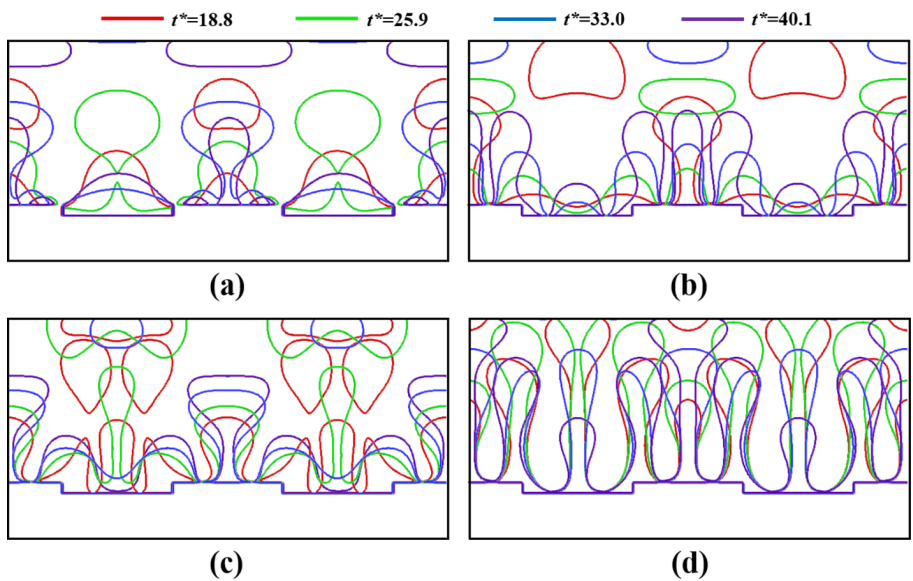

FIG. 14. The interfacial evolution of the heating surface in the CHF state: (a) $E_0 = 0$, (b) $E_0 = 0.050$, (c) $E_0 = 0.065$, and (d) $E_0 = 0.080$.

## V. CONCLUSIONS

In this study, lattice Boltzmann simulations find that coupled electric fields and microcavities with conducting–insulating surfaces can enhance boiling heat transfer synergistically. When the microcavity corners are insulated, the electrical forces can be reorganized, which improves the bubble formation process. The heating surface's boiling characteristics and heat transfer mechanism were analyzed under different electric field intensities. The main conclusions are as follows:

(1) Conducting–insulating surfaces under the action of the electric field can eliminate the field trap effect caused by the microcavity structure. After insulating the corner structure, the bottom and the top of the microcavity generate a large electrical force, rapidly pushing the bubbles out of the heating surface and avoiding the mutual merging of the bubbles.
(2) The electrical force can achieve CHF enhancement by destabilizing the vapor block. When the electric field intensity is low, intermittent breakup and regeneration of the vapor block in the microcavity occurs under the perturbation of the electrical force. When the electric field intensity is large, the electrical force pins the contact line at the junction of the conducting and insulating surfaces, which facilitates the formation of multiple continuously open vapor–liquid separation paths, thereby improving the boiling heat transfer performance of the microcavity and significantly enhancing the CHF.
(3) The prediction of CHF on microcavity surfaces under the action of electric fields requires consideration of structural effects. At the junction of the conducting and insulating surfaces, the violent twisting of the iso-potential lines produces a high electric field intensity. The CHF correlation equation for the non-uniform electric field can be established by modifying the electric field intensity.

## ACKNOWLEDGMENTS

The authors are grateful for the support of the National Natural Science Foundation of China (Grant Nos. 11872083 and 12172017).

## AUTHOR DECLARATIONS
### Conflict of Interest

The authors have no conflicts to disclose.

### Author Contributions

**Fanming Cai:** Conceptualization (equal); Investigation (equal); Methodology (equal); Validation (equal); Writing – original draft (equal); Writing – review & editing (equal). **Zhaomiao Liu:** Formal analysis (equal); Investigation (equal); Resources (equal); Software (equal); Writing – review & editing (equal). **Nan Zheng:** Software (equal); Validation (equal). **Yan Pang:** Resources (equal); Software (equal); Writing – original draft (equal).

## DATA AVAILABILITY

The data that support the findings of this study are available from the corresponding author upon reasonable request.